\documentclass[12pt]{article}

\usepackage{amssymb}
\usepackage{amsmath}
\usepackage[pdftex]{graphicx}   %For arXiv
\usepackage{amsthm}   %This is necessary for "proof"
\usepackage{authblk}   %This is necessary for author descriptions
\usepackage{lscape}   %This is necessary for "landscape"
\usepackage{comment}
\usepackage{setspace}
\usepackage{multirow}
\usepackage{url}
\usepackage[super]{cite}
\makeatletter

\@addtoreset{equation}{section}
\makeatother

\newtheorem{defi}{Definition}
\newtheorem{theo}{Theorem}

\makeatletter
\def\mojiparline#1{
\newcounter{mpl}
\setcounter{mpl}{#1}
\@tempdima=\linewidth
\advance\@tempdima by-\value{mpl}zw
\addtocounter{mpl}{-1}
\divide\@tempdima by \value{mpl}
\advance\kanjiskip by\@tempdima
\advance\parindent by\@tempdima
}
\renewcommand{\@biblabel}[1]{#1.}
\makeatother

\allowdisplaybreaks[4]

\newcommand{\indep}{\mathop{\perp\!\!\!\perp}}

\newcommand{\bld}{\boldsymbol}

\usepackage[top=1in,bottom=1in,left=1in,right=1in]{geometry}

\title{Evaluating the Conservativeness of Robust Sandwich Variance Estimator in Weighted Average Treatment Effects}
\author[1]{Shunichiro Orihara \thanks{Corresponding author\\ \hspace{0.4cm} Address: 6-1-1 Shinjuku, Shinjuku-ku, Tokyo 160-8402, Japan \\ \hspace{0.4cm} Email: orihara@tokyo-med.ac.jp}}
\affil[1]{Department of Health Data Science, Tokyo Medical University, Tokyo, Japan}
\date{}

\begin{document}
\begin{singlespace}
\maketitle
\end{singlespace}
\section*{Abstract}
In causal inference, the Inverse Probability Weighting (IPW) estimator is commonly used to estimate causal effects for estimands within the class of Weighted Average Treatment Effect (WATE). When constructing confidence intervals (CIs), robust sandwich variance estimators are frequently used for practical reasons. Although these estimators are easy to calculate using widely-used statistical software, they often yield narrow CIs for commonly applied estimands, such as the Average Treatment Effect on the Treated and the Average Treatment Effect for the Overlap Populations. In this manuscript, we reexamine the asymptotic variance of the IPW estimator and clarify the conditions under which CIs derived from the sandwich variance estimator are conservative. Additionally, we propose new criteria to assess the conservativeness of CIs. The results of this investigation are validated through simulation experiments and real data analysis.

\vspace{0.5cm}
\noindent
{\bf Keywords}: Balancing weight, Causal estimands, Causal null hypothesis, Inverse probability weighting, Propensity score\\

\section{Introduction} %%580 words
In causal inference, selecting a valid causal estimand that represents causal effect interpretation is a crucial step in addressing clinically questions. The Average Treatment Effect (ATE) is a widely recognized causal estimand, comparing outcomes if all subjects receive an active treatment versus if they all receive a control treatment. The Average Treatment Effect on the Treated (ATT) is also well considered, comparing outcomes if treated subjects retain their treatment versus if they switch to a control treatment. Recently, the Average Treatment Effects for the Overlap Population (ATO), an estimand proposed by Li et al.\cite{Li2018}, has gained attention. It offers several advantages over the ATE. For instance, the ATO assigns larger weights to subjects more likely to switch from their actual treatment to the `counterfactual' treatment, making it a more robust measure of causal effect. The estimands belong in the category of `Weighted Average Treatment Effect' (WATE\cite{Hi2003,Li2018}), as they are viewed as specific weighted versions of the ATE. Further details can be found in Li et al.\cite{Li2018}.

The estimands can be estimated as a weighted average of the outcome defined by the (estimated) propensity score \cite{Ro1983}. The well-known estimator is the Inverse Probability (Treatment) Weighting (IP(T)W) estimator for the ATE. Since other weighted estimators for the WATE can be constructed in the same manner, we refer to these weighted estimators as the `IPW estimator' in this manuscript\cite{Ma2019}. The confidence interval (CI) for the IPW estimator is commonly derived from the asymptotic variance, taking into account estimating equations for both the weighted average of the outcome and the propensity score. However, in practice, the CI is often derived based only on the former; the uncertainty of the propensity score is sometimes overlooked. This is because implementing the `robust sandwich variance' estimator for the weighted average of the outcome is straightforward. For example, {\it sandwich} in R, or the {\it WHITE} option in {\it REG} procedure in SAS can be easily implemented \cite{Re2022}.

For the IPW estimator of the ATE, it is well-known that a CI that ignores the uncertainty of the propensity score tends to yield a more conservative CI compared to one that accounts for it\cite{Lu2004}. We refer to the former as the `simple CI' and the latter as the `exact CI' in this manuscript. However, recent findings suggest that the simple CI for the ATT estimator might not always be conservative\cite{Re2022}. Consequently, the exact CI is thought of as more suitable for the IPW estimator for the ATT. From a practical standpoint, it is important to identify specific conditions under which the simple CI yields a conservative CI. Unfortunately, Reifeis and Hudgens \cite{Re2022} did not describe these situations clearly.

In this manuscript, we confirm again the asymptotic variance of the IPW estimator and clarify the conditions under which the simple CI yields a conservative CI. We focus on binary outcomes, which are commonly encountered in epidemiologic studies. In Section 2, we detail the form of the asymptotic variance for the IPW estimator within the WATE class, extending the works of Mao et al.\cite{Ma2019} and Reifeis and Hudgens \cite{Re2022}. Section 3 investigates scenarios where the simple CI is conservative. Our findings show that under the Fisher sharp null hypothesis\cite{Im2015}, or concerning the conditional ATE\cite{Cr2006}, the simple CI is accurately conservative, thereby reducing concerns about $\alpha$-error inflation. Moreover, we explore broader situations and propose new criteria to assess whether the simple CI remains conservative. Then, validate these findings through simple simulation settings and a real-world data example on bladder cancer\cite{Sh2017}.

\section{Methods} %370 words
\subsection{Definition of the WATE}
Let $n$ be the sample size. $T_{i}\in\{0,1\}$, $\boldsymbol{X}_{i}\in\mathcal{X}\subset\mathbb{R}^{p}$ and $(Y_{1i},Y_{0i})$ represent the treatment, a vector of covariates measured prior to treatment, and potential outcomes, respectively. In the main manuscript, we consider the binary outcome; $Y_{ti}\in\{0,1\}$. Based on the stable unit treatment value assumption\cite{Ro1983}, the observed outcome is defined as $Y_{i}:=T_{i}Y_{1i}+(1-T_{i})Y_{0i}$. Under these settings, we assume that i.i.d.\ copies $(T_{i},\bld{X}_{i},Y_{i})$, $i=1,\, 2,\, \dots,\, n$ are obtained. We further assume the strongly ignorable treatment assignment $(Y_{1i},Y_{0i})\indep T_{i}|\bld{X_{i}}$\cite{Ro1983} for the subsequent discussions.

We now introduce the WATE. The ATE conditional on $\bld{x}$ is defined as $\tau(\bld{x}):={\rm E}[Y_{1}-Y_{0}|\bld{x}]$, and the WATE is defined as
$$
\tau_{w}=\mu_{w1}-\mu_{w0}:=\frac{{\rm E}[\tau(\bld{X})w(e(\bld{X}))]}{{\rm E}[w(e(\bld{X}))]},
$$
where $e\equiv e(\bld{X}):={\rm Pr}\left(T=1|\bld{X}\right)$ is the propensity score\cite{Ro1983}, and $w(\cdot)$ represents the weight function for $e$. When $w(e)\equiv 1$, the WATE becomes the ATE: $\tau_{ATE}:={\rm E}[Y_{1}-Y_{0}]$. When $w(e)=e$, the WATE becomes the ATT
\begin{align*}
\tau_{ATT}:=\frac{{\rm E}[\tau(\bld{X})e(\bld{X})]}{{\rm E}[e(\bld{X})]}={\rm E}[\tau(\bld{X})|T=1]={\rm E}[Y_{1}-Y_{0}|T=1].
\end{align*}
When $w(e)=e(1-e)$, the WATE becomes the ATO\cite{Li2018}
\begin{align*}
\tau_{ATO}:=\frac{{\rm E}[\tau(\bld{X})e(\bld{X})(1-e(\bld{X}))]}{{\rm E}[e(\bld{X})(1-e(\bld{X}))]}.
\end{align*}
Given that the function $e(1-e)$, with $e\in(0,1)$, is convex and symmetric around $0.5$, subjects with a propensity score close to $0.5$ (meaning they could easily change their actual treatment to the ``counterfactual'' treatment) are weighted more heavily than those near $0$ or $1$.

\subsection{IPW Estimator and the Asymptotic Property of the WATE}
As discussed in Mao et al.\cite{Ma2019}, the WATE can be estimated using the IPW estimator
\begin{align*}
\hat{\tau}_{w}=\hat{\mu}_{w1}-\hat{\mu}_{w0}=\frac{\sum_{i=1}^{n}W_{i}T_{i}Y_{i}}{\sum_{i=1}^{n}W_{i}T_{i}}-\frac{\sum_{i=1}^{n}W_{i}(1-T_{i})Y_{i}}{\sum_{i=1}^{n}W_{i}(1-T_{i})},
\end{align*}
where
$$
W_{i}=\frac{w(\hat{e}_{i})}{T_{i}\hat{e}_{i}+(1-T_{i})(1-\hat{e}_{i})},
$$
and $\hat{e}_{i}\equiv e_{i}(\hat{\bld{\alpha}})=expit\left\{\bld{X}_{i}^{\top}\hat{\bld{\alpha}}\right\}$ denotes the estimated propensity score. In this manuscript, $\hat{\bld{\alpha}}$ is a Maximum Likelihood Estimator (MLE). Under the settings, the asymptotic distribution of the IPW estimator becomes
$$
\sqrt{n}\left(\hat{\tau}_{w}-\tau_{w}^{0}\right)\stackrel{L}{\to}N\left(\bld{0},\sigma^2\right),
$$
where the asymptotic variance $\sigma^2$ is described as
\begin{align}
\label{asy2}
\sigma^{2}=\frac{1}{a_{22}^2}(1,-1)\left(B_{22}+\delta A_{11}^{-1}\delta^{\top}-B_{12}A_{11}^{-1}B_{12}^{\top}\right)\left(
\begin{array}{c}
1\\
-1
\end{array}
\right),
\end{align}
with
$$
A_{11}={\rm E}\left[e(1-e)\bld{X}^{\otimes 2}\right],\ \ a_{22}={\rm E}[w(e)],
$$
$$
B_{12}=\left(
\begin{array}{c}
b_{1}^{\top}\\
b_{2}^{\top}
\end{array}
\right):=\left(
\begin{array}{c}
{\rm E}\left[w(e)(Y_{1}-\mu_{w1})(1-e)\bld{X}^{\top}\right]\\
-{\rm E}\left[w(e)(Y_{0}-\mu_{w0})e\bld{X}^{\top}\right]
\end{array}
\right),
$$
$$
B_{22}=\left(
\begin{array}{cc}
{\rm E}\left[\frac{w(e)^2(Y_{1}-\mu_{w1})^2}{e}\right]&\\
0&{\rm E}\left[\frac{w(e)^2(Y_{0}-\mu_{w0})^2}{1-e}\right]
\end{array}
\right),
$$
and
$$
\delta=\left(
\begin{array}{c}
\delta_{1}^{\top}\\
\delta_{2}^{\top}
\end{array}
\right):=
\left(
\begin{array}{c}
{\rm E}\left[w'(e)(Y_{1}-\mu_{w1})e(1-e)\bld{X^{\top}}\right]\\
{\rm E}\left[w'(e)(Y_{0}-\mu_{w0})e(1-e)\bld{X^{\top}}\right]
\end{array}
\right).
$$
For more details on the mathematical computations, please see Appendix A.

The first term of (\ref{asy2}) forms the primary component of the asymptotic variance of the IPW estimator. Standard sandwich variance calculation functions, such as {\it sandwich} in R, compute only this term. The second and third terms of (\ref{asy2}) are related to the variability of the propensity score estimation and are typically ignored when using the sandwich variance estimator. The CI constructed from the first term of (\ref{asy2}) is referred to as the `simple CI', whereas the CI constructed from all terms of (\ref{asy2}) is termed the `exact CI' in this manuscript.

\subsection{Definition of Null Hypotheses and Homogeneous Effects}
In this manuscript, some null hypotheses are considered. The Fisher sharp null, the Neyman null\cite{Im2015}, and the null hypothesis concerning the conditional ATE\cite{Cr2006} represent as follows.
\begin{defi}{Fisher sharp null hypothesis (SN)}
$$
Y_{i1}=Y_{i0},\ \ \ \ \ ^\forall  i=1,2,\dots,n
$$
\end{defi}
\begin{defi}{Neyman null hypothesis (NN)}
$$
\tau_{w}=0\ \ \Leftrightarrow\ \ \frac{{\rm E}[w(e(\bld{X})){\rm E}\left[Y_{1}|\bld{X}\right]]}{{\rm E}[w(e(\bld{X}))]}=\frac{{\rm E}[w(e(\bld{X})){\rm E}\left[Y_{0}|\bld{X}\right]]}{{\rm E}[w(e(\bld{X}))]}
$$
\end{defi}
\begin{defi}{Null hypothesis concerning the conditional ATE (CN)}
$$
\tau(\bld{x})=0,\ \ \ \ \ ^\forall \bld{x}\in\mathcal{X}
$$
\end{defi}\noindent
From the definitions, $SN\Rightarrow CN \Rightarrow NN$.

Additionally, we define the homogeneous treatment effects as the complement of treatment effects heterogeneity\cite{Wa2018}.
\begin{defi}{Homogeneous treatment effects}
$$
\tau(\bld{x})=\tau,\ \ \ \ \ ^\forall \bld{x}\in\mathcal{X}
$$
\end{defi}\noindent
Under the homogeneous treatment effects, $\tau_{w}\equiv\tau_{ATE}$ for all $w(\cdot)$. Also, $NN$ becomes equivalent to $CN$.

\section{Results} % 660 words
\subsection{Properties of the Simple CIs} 
In this section, we examine scenarios where simple CIs yield conservative results. In other words, we clarify the conditions under which the second and third terms of (\ref{asy2}) can be ignored, thus preventing misinterpretation of statistical analysis results in practical applications.
\subsubsection{For the ATE}
From (\ref{asy2}), a well-known conclusion can be simply derived.
\begin{theo}$\phantom{a}$\\
For the ATE, the weight function is given by $w(e)\equiv1$. Therefore, $w'(e)=0$. This imply that $\delta=O$, and the second and third term of (\ref{asy2}) become precisely negative.
\end{theo}\noindent
This theorem indicates that the simple CI always yields a conservative CI when we are interested in the ATE.

\subsubsection{For the ATT}
As mentioned by Reifeis and Hudgens \cite{Re2022}, from the form of (\ref{asy2}), it isn't clear whether the standard sandwich variance for ATT, ATO, or certain estimands is conservative. From this point onward, to understand the asymptotic variance more clearly, we calculate the second and third term of (\ref{asy2}) more precisely.

The values of $\delta$ and $B_{22}$ become the following, respectively:
\begin{align*}
\delta&=\left(
\begin{array}{c}
\delta_{1}^{\top}\\
\delta_{2}^{\top}
\end{array}
\right)=\left(
\begin{array}{c}
{\rm E}\left[(Y_{1}-\mu_{w1})e(1-e)\bld{X^{\top}}\right]\\
{\rm E}\left[(Y_{0}-\mu_{w0})e(1-e)\bld{X^{\top}}\right]
\end{array}
\right),\\
B_{12}&=\left(
\begin{array}{c}
b_{1}^{\top}\\
b_{2}^{\top}
\end{array}
\right)=\left(
\begin{array}{c}
{\rm E}\left[(Y_{1}-\mu_{w1})e(1-e)\bld{X}^{\top}\right]\\
-{\rm E}\left[(Y_{0}-\mu_{w0})e^2\bld{X}^{\top}\right]
\end{array}
\right)=\left(
\begin{array}{c}
\delta_{1}^{\top}\\
\delta_{2}^{\top}-\alpha^{\top}
\end{array}
\right),
\end{align*}
where $\alpha={\rm E}\left[(Y_{0}-\mu_{w0})e\bld{X}\right]$. Therefore, the second and third term of (\ref{asy2}) become
\begin{align}
\label{ATTsnd}
(1,-1)\left(\delta A_{11}^{-1}\delta^{\top}-B_{12}A_{11}^{-1}B_{12}^{\top}\right)\left(
\begin{array}{c}
1\\
-1
\end{array}
\right)&\nonumber\\
&\hspace{-6.5cm}=\delta_{1}^{\top}A_{11}^{-1}\delta_{1}-2\delta_{1}^{\top}A_{11}^{-1}\delta_{2}+\delta_{2}^{\top}A_{11}^{-1}\delta_{2}-b_{1}^{\top}A_{11}^{-1}b_{1}+2b_{1}^{\top}A_{11}^{-1}b_{2}-b_{2}^{\top}A_{11}^{-1}b_{2}\nonumber\\
&\hspace{-6.5cm}=-\alpha^{\top}A_{11}^{-1}\alpha+2(\delta_{2}-\delta_{1})^{\top}A_{11}^{-1}\alpha
\end{align}
From the results, the following conclusions can be derived.
\begin{theo}$\phantom{a}$\\
For the ATT, when $SN$, $CN$, or the treatment effects are homogeneous, the second and third terms of (\ref{asy2}) are precisely negative. Outside of these situations, when covariates have the positive support: $\mathcal{X}\subset\mathbb{R}^{p}_{\geq0}$, a sufficient condition for the second and third terms of (\ref{asy2}) to be precisely negative is that the following inequality holds:
\begin{align}
\label{ineq_att}
\gamma{\rm E}\left[e(1-e)\bld{X}^{\top}\right]{\rm E}\left[e(1-e)\bld{X}^{\otimes 2}\right]^{-1}{\rm E}\left[e(1-e)\bld{X}\right]&\nonumber\\
&\hspace{-7.5cm}\leq{\rm E}\left[e(Y_{0}-\mu_{w0})\bld{X}^{\top}\right]{\rm E}\left[e(1-e)\bld{X}^{\otimes 2}\right]^{-1}{\rm E}\left[e(Y_{0}-\mu_{w0})\bld{X}\right].
\end{align}
When considering $NN$, $\gamma$ is set to 4. Otherwise, $\gamma$ is set to 16.
\end{theo}\noindent
The proof of Theorem 2 is in Appendix B. Note that while the former statement can be proven in the broader context of WATE, not limited to ATT, ATO, or binary outcome situations, we focus solely on these specific cases to simplify the proof.

This theorem suggests that the simple CI yields a conservative CI when focusing on certain null hypotheses or in the presence of a homogeneous treatment effect. Moreover, since the right-hand side of (\ref{ineq_att}) is positive, we propose a new criterion to determine whether the simple CI is conservative: by verifying whether the left-hand side of (\ref{ineq_att}) is approximately $0$. If the left-hand side of (\ref{ineq_att}) is sufficiently close to $0$, it is expected that the inequality (\ref{ineq_att}) will be satisfied.

\subsubsection{For the ATO}
Next, we consider the asymptotic variance of the ATO. In the same manner as the ATT, the values of $\delta$ and $B_{22}$ become the following, respectively:
\begin{align*}
\delta&=\left(
\begin{array}{c}
\delta_{1}^{\top}\\
\delta_{2}^{\top}
\end{array}
\right)=\left(
\begin{array}{c}
{\rm E}\left[(Y_{1}-\mu_{w1})e(1-e)(1-2e)\bld{X^{\top}}\right]\\
{\rm E}\left[(Y_{0}-\mu_{w0})e(1-e)(1-2e)\bld{X^{\top}}\right]
\end{array}
\right),\\
B_{12}&=\left(
\begin{array}{c}
b_{1}^{\top}\\
b_{2}^{\top}
\end{array}
\right)=\left(
\begin{array}{c}
{\rm E}\left[(Y_{1}-\mu_{w1})e(1-e)^2\bld{X}^{\top}\right]\\
-{\rm E}\left[(Y_{0}-\mu_{w0})e^2(1-e)\bld{X}^{\top}\right]
\end{array}
\right)=\left(
\begin{array}{c}
\delta_{1}^{\top}+\alpha_{1}^{\top}\\
\delta_{2}^{\top}-\alpha_{2}^{\top}
\end{array}
\right),
\end{align*}
where $\alpha_{1}={\rm E}\left[(Y_{1}-\mu_{w1})e^{2}(1-e)\bld{X}\right]$ and $\alpha_{2}={\rm E}\left[(Y_{0}-\mu_{w0})e(1-e)^{2}\bld{X}\right]$. Through the similar calculation as the ATT (\ref{ATTsnd}), the second and third term of (\ref{asy2}) become
$$
(1,-1)\left(\delta A_{11}^{-1}\delta^{\top}-B_{12}A_{11}^{-1}B_{12}^{\top}\right)\left(
\begin{array}{c}
1\\
-1
\end{array}
\right)=-(\alpha_{1}+\alpha_{2})^{\top}A_{11}^{-1}(\alpha_{1}+\alpha_{2})+2(\delta_{2}-\delta_{1})^{\top}A_{11}^{-1}(\alpha_{1}+\alpha_{2})\nonumber\\
$$
From the results, the following conclusions can be derived.
\begin{theo}$\phantom{a}$\\
For the ATO, when $SN$, $CN$, or the treatment effects are homogeneous, the second and third terms of (\ref{asy2}) are precisely negative. Outside of these situations, when covariates have the positive support: $\mathcal{X}\subset\mathbb{R}^{p}_{\geq0}$, a sufficient condition for the second and third terms of (\ref{asy2}) to be precisely negative is that the following inequality holds:
\begin{align}
\label{ineq_ato}
\gamma{\rm E}\left[e(1-e)(1-2e)\bld{X}^{\top}\right]{\rm E}\left[e(1-e)\bld{X}^{\otimes 2}\right]^{-1}{\rm E}\left[e(1-e)(1-2e)\bld{X}\right]&\nonumber\\
&\hspace{-10.5cm}\leq{\rm E}\left[e(1-e)\left\{e(Y_{1}-\mu_{w1})+(1-e)(Y_{0}-\mu_{w0})\right\}\bld{X}^{\top}\right]{\rm E}\left[e(1-e)\bld{X}^{\otimes 2}\right]^{-1}\nonumber\\
&\hspace{-10cm}\times {\rm E}\left[e(1-e)\left\{e(Y_{1}-\mu_{w1})+(1-e)(Y_{0}-\mu_{w0})\right\}\bld{X}\right].
\end{align}
When considering $NN$, $\gamma$ is set to 4. Otherwise, $\gamma$ is set to 16.
\end{theo}\noindent
Since the proof follows the same procedure as for the ATT situation, we omit it here.

Following the same discussion as with the ATT situation, we also propose a new criterion to determine whether the simple CI is conservative: by verifying whether the left-hand side of (\ref{ineq_ato}) is approximately $0$. If the left-hand side of (\ref{ineq_ato}) is sufficiently close to $0$, it is expected that the inequality (\ref{ineq_ato}) will be satisfied.

\subsection{Comparison of the Proposed Criteria}
\begin{figure}[h]
\begin{center}
\includegraphics[width=12cm]{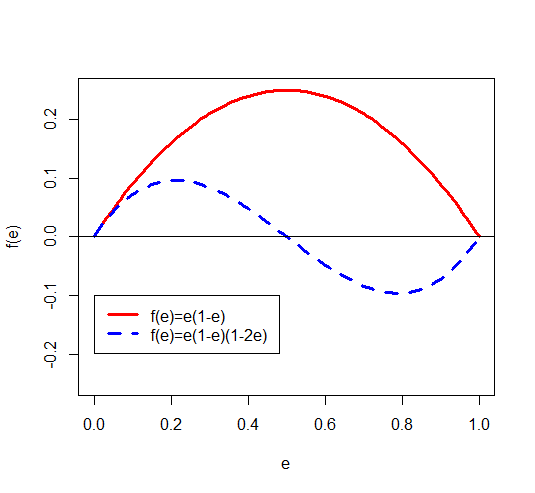}
\caption{Function forms of the component of the proposed criteria}
\label{fig1}
\end{center}
\end{figure}

We compare the proposed criteria to determine whether the simple CI yields a conservative CI: assessing if the left-hand side of (\ref{ineq_att}) or (\ref{ineq_ato}) is sufficiently close to $0$. Given that ${\rm E}\left[e(1-e)\boldsymbol{X}^{\otimes 2}\right] > O$, it is essential to satisfy to assess whether ${\rm E}\left[e(1-e)\bld{X}\right]\approx\bld{0}$ for the ATT or ${\rm E}\left[e(1-e)(1-2e)\bld{X}\right]\approx\bld{0}$ for the ATO. To evaluate these conditions, we consider the functions $f(e)=e(1-e)$ and $f(e)=e(1-e)(1-2e)$ for $e\in(0,1)$. As shown in Figure \ref{fig1}, the function for the ATT forms a parabolic shape with a maximum value at 0.5, while the function for the ATO resembles a sine-like curve, varying between positive and negative around 0.5. This suggests that meeting the condition for the ATT is more challenging, whereas the condition for the ATO is likely easier to satisfy. In other words, the simple CI for the ATO tends to be more conservative compared to that for the ATT. To further explore these properties, we conduct simulation experiments in the following section.

\subsection{Simulation Experiments} %% 330 words
In this section, we demonstrate that under CN conditions, the simple CI for the IPW estimator of both ATT and ATO yields conservative results. Furthermore, we illustrate that the proposed criteria can effectively determine whether the simple CI is conservative under NN conditions.

\subsubsection{Data-Generating Mechanism}
First, we will describe the Data-Generating Mechanism (DGM) used in this simulation. We consider two confounders, denoted as $X_{i1} \stackrel{i.i.d.}{\sim} N(50, 5^2)$ and $X_{i2} \stackrel{i.i.d.}{\sim} \text{Bin}(0.5)$. The assignment mechanism (true propensity score) related to these confounders is defined by ${\rm Pr}\left(T=1|\bld{X}_{i}\right)=expit\left\{\varepsilon\times(0.03(X_{i1}-50)+0.3(X_{i2}-0.5))\right\}$. Here, $\varepsilon$ adjusts the relationship between the assignment and the confounders, influencing the variance of the propensity score. The outcome model is specified as ${\rm Pr}\left(Y_{0,i}=1|\bld{X}_{i}\right)=expit\left\{(1,X_{i1}-50,X_{i2}-0.5)\bld{\beta}_{0}\right\}$ and ${\rm Pr}\left(Y_{1,i}=1|\bld{X}_{i}\right)=expit\left\{(1,X_{i1}-50,X_{i2}-0.5)\bld{\beta}_{1}\right\}$.

In subsequent simulation experiments, the parameters $\varepsilon$, $\bld{\beta}_{0}$, and $\bld{\beta}_{1}$ will be varied. Under CN conditions, we consider two scenarios; \#1: $\varepsilon=1$ and $\bld{\beta}_{0}=\bld{\beta}_{1}=(0,-0.2,0.1)^{\top}$, and \#2: $\varepsilon=3$ and $\bld{\beta}_{0}=\bld{\beta}_{1}=(0,-0.2,0.1)^{\top}$. Under NN conditions, we also consider two scenarios; \#3: $\varepsilon=1$, $\bld{\beta}_{0}=(0,-0.25,0.1)^{\top}$, and $\bld{\beta}_{1}=(0.01,-0.4,4)^{\top}$, and \#4: $\varepsilon=3$, $\bld{\beta}_{0}=(1,-0.5,-0.5)^{\top}$, and $\bld{\beta}_{1}=(1,0.5,0.5)^{\top}$. In the former scenarios, the CN condition is exactly satisfied, whereas in the latter scenarios, the NN condition is approximately met.

\subsubsection{Performance Metrics}
Simulation experiments are assessed by calculating the mean, empirical standard error (ESE), and the $\alpha$-error across 1000 iterations. The $\alpha$-error is calculated using the simple CI with both the true and estimated propensity scores, and using the exact CI with only the estimated propensity score. We then determine the proportion of cases in which the simple CI and the exact CI does not include 0.

\subsubsection{Simulation Results}
All simulation results are summarized in Table \ref{tab1} and \ref{tab2}. Under the CN condition, both ATT and ATO exhibit conservative simple CIs, compared with the exact CIs. This tendency also holds when the variance of the propensity score is large ($\varepsilon=3$).

Under the NN condition, with scenario \#3, the left-hand side of (\ref{ineq_ato}) is approximately $1.07 \times 10^{-2}$, while the right-hand side is $1.84 \times 10^{-2}$. In this scenario, the simple CI yields a conservative result. However, when scenario \#4, the left-hand side of (\ref{ineq_ato}) is approximately $7.04 \times 10^{-2}$, in contrast to the right-hand side, which is $0.09 \times 10^{-2}$. This indicates that the proposed criterion is not satisfied. In such cases, the simple CI exhibits $\alpha$-error inflation (5.5 vs. 4.9).

\begin{table}[h]
\begin{center}
\caption{Summary Statistics Under the Null Hypothesis Concerning the Conditional ATE (CN): The sample size is 2000, and the number of iterations is 1000. The $\alpha$-errors are calculated using both the simple CI, which ignores the variability of the estimated propensity score, and the exact CI, which considers the variability.}
\begin{tabular}{ccc|cccc}\hline
&&&\multicolumn{4}{|c}{\bf Summary statistics}\\
&&{\bf Propensity}&{\bf Mean}&{\bf ESE}&\multicolumn{2}{c}{\bf $\alpha$-error:}\\
{\bf Scenario \#}&{\bf Estimand}&{\bf score}&{\footnotesize ($\times 10^{-2}$)}&{\footnotesize ($\times 10^{-2}$)}&{\bf simple CI} {\footnotesize (\%)}&{\bf exact CI} {\footnotesize (\%)}\\\hline\hline
1&ATT&True&0.080&2.224&4.1&--\\\cline{3-7}
&&Estimated&0.094&2.036&2.9&4.4\\\cline{2-7}
&ATO&True&0.068&2.210&4.4&--\\\cline{3-7}
&&Estimated&0.085&2.018&2.3&4.4\\\hline
2&ATT&True&0.086&2.418&4.7&--\\\cline{3-7}
&&Estimated&0.119&2.223&3.2&4.5\\\cline{2-7}
&ATO&True&0.053&2.313&4.2&--\\\cline{3-7}
&&Estimated&0.084&2.101&2.5&4.4\\\hline
\end{tabular}
\label{tab1}
\end{center}
{\footnotesize
ESE: empirical standard error; CI: confidence interval (calculated as 95\%CI)\\
Situation \#1: $\varepsilon=1$ and $\bld{\beta}_{0}=\bld{\beta}_{1}=(0,-0.2,0.1)^{\top}$\\
Situation \#2: $\varepsilon=3$ and $\bld{\beta}_{0}=\bld{\beta}_{1}=(0,-0.2,0.1)^{\top}$
}
\end{table}

\begin{table}[h]
\begin{center}
\caption{Summary Statistics Under the Neyman Null (NN) for the ATO: The sample size is 2000, and the number of iterations is 1000. The $\alpha$-errors are calculated using both the simple CI, which ignores the variability of the estimated propensity score, and the exact CI, which considers the variability.}
\begin{tabular}{cc|cccc}\hline
&&\multicolumn{4}{|c}{\bf Summary statistics}\\
&{\bf Propensity}&{\bf Mean}&{\bf ESE}&\multicolumn{2}{c}{\bf $\alpha$-error:}\\
{\bf Scenario \#}&{\bf score}&{\footnotesize ($\times 10^{-2}$)}&{\footnotesize ($\times 10^{-2}$)}&{\bf simple CI} {\footnotesize (\%)}&{\bf exact CI} {\footnotesize (\%)}\\\hline\hline
3&True&0.162&2.250&5.3&--\\\cline{2-6}
&Estimated&0.160&1.878&2.4&4.8\\\hline
4&True&-0.045&2.383&5.8&--\\\cline{2-6}
&Estimated&-0.049&2.428&5.5&4.9\\\hline
\end{tabular}
\label{tab2}
\end{center}
{\footnotesize
ESE: empirical standard error; CI: 95\% confidence interval (calculated as 95\%CI)\\
Situation \#3: $\varepsilon=1$, $\bld{\beta}_{0}=(0,-0.25,0.1)^{\top}$, and $\bld{\beta}_{1}=(0.01,-0.4,4)^{\top}$\\
Situation \#4: $\varepsilon=3$, $\bld{\beta}_{0}=(1,-0.5,-0.5)^{\top}$, and $\bld{\beta}_{1}=(1,0.5,0.5)^{\top}$
}
\end{table}

\subsection{Illustrative Real Data Example} %%180 words
In this section, we provide an illustrative example of a real data analysis conducted on muscle-invasive bladder cancer\cite{Sh2017}. Specifically, we examined the causal effects of adjuvant therapies on muscle-invasive bladder cancer-related deaths. We considered cancer-specific death as a binary endpoint, which differed from the outcome in the original manuscript, that is, we disregarded the `time information' from the survival data. The dataset consisted of 322 patients: 74 in the adjuvant chemotherapy (treatment) group and 248 in the radical cystectomy (control) group. Cancer-specific deaths occurred in 23 (31.1 \%) and 45 patients (18.2 \%) in the treatment and control groups, respectively. For more information on this study, please refer to Shimizu et al.\cite{Sh2017}.

All analysis results are summarized in Table \ref{tab3}. Given that the proposed criteria under NN for the ATT and ATO are 0.535 and 0.125, respectively, it is expected that the simple CI will not yield a conservative CI. In this analysis, we assume interest in either SN or CN conditions. Under this assumption, the null hypotheses, which propose no causal effects, are not rejected using the simple CI.

\begin{table}[h]
\begin{center}
\caption{Summary Statistics of Real Data Analysis: The CIs are calculated using both the simple CI, which ignores the variability of the estimated propensity score, and the exact CI, which considers the variability.}
\begin{tabular}{c|ccc}\hline
&\multicolumn{3}{|c}{\bf Summary statistics}\\
{\bf Estimand}&{\bf Point Estimate}&{\bf 95\%CI} ({\footnotesize simple CI})&{\bf 95\%CI} ({\footnotesize exact CI})\\\hline\hline
ATT&-0.0478&(-0.2004, 0.1049)&(-0.2001, 0.1045)\\\hline
ATO&0.0017&(-0.1394, 0.1428)&(-0.1392, 0.1426)\\\hline
\end{tabular}
\label{tab3}
\end{center}
{\footnotesize
CI: confidence interval
}
\end{table}

\section{Discussion} %% 580 words
In this manuscript, we investigate scenarios where a confidence interval that ignores the variability of the estimated propensity score (referred to as `simple CI') for the IPW estimator is conservative compared to one that considers this variability (termed `exact CI') in weighted ATE classes. This consideration is important from a practical standpoint, as the sandwich variance, which is commonly used to derive the simple CI, is widely employed. Specifically, under the Fisher sharp null, conditional ATE null, or homogeneous treatment effect scenarios, the simple CI yields a conservative CI. However, in other scenarios, by evaluating terms related to the variability of the estimated propensity score in the asymptotic variance of the IPW estimator, the behavior of the simple CI can be forecasted. Using this property, we propose simple criteria to assess whether the simple CI is conservative. The results of this investigation are validated through simulation experiments and real data analysis.

To the best of our knowledge, this is the first study to identify situations where the simple CI becomes conservative. It is well known that the sandwich variance of the IPW estimator for the ATE is conservative, whereas for the ATT, it is not\cite{Re2022}. However, from a practical standpoint, the sandwich variance is commonly applied to the Weighted Average Treatment Effect estimands, including the ATT and ATO\cite{Me2021,Co2022}. According to the results of this manuscript, statistical hypothesis tests considering certain causal null hypotheses are appropriate. In other scenarios, it is indeed necessary to address this issue by using the exact CI or the proposed criteria.

The results of the simulation experiments suggest that the proposed criteria, especially for the ATO, are effective in determining the behavior of the simple CI. However, it should be noted that the right-hand side of (\ref{ineq_ato}) tends to yield a small value. Therefore, additional caution is needed when practically applying these criteria. Specifically, we believe the following two points are important: 1) the distribution of the estimated propensity score, and 2) its variance. As for the first point, this is evident from Figure \ref{fig1}. The blue dashed line indicates that if the distribution is symmetrical around 0.5, the criterion is likely to be satisfied. Regarding the second point, a large variance in the estimated propensity score can lead to a significant difference in the value of ${\rm E}\left[e(1-e)(1-2e)\boldsymbol{X}\right]$, which is the fundamental term of the criterion for the ATO. Conversely, if the variance is small, this term tends to yield similar values since the estimated propensity scores are concentrated around 0.5. This trend is observed in simulation scenarios \#3 and \#4. In summary, when the estimated propensity score distribution is centered around 0.5, the criterion for the ATO is more likely to be effective.

In future work, it will be important to examine the behavior of other estimands, such as matching weight, entropy weight, and others\cite{Li2013,Ma2020}. For the matching weight, it may be necessary to employ approximation techniques, as this weight includes the non-differentiable point\cite{Li2013,Or2022}. Additionally, while some considerations for continuous outcomes are presented in Appendix C, addressing time-to-event outcomes is also crucial. Mao et al.\cite{Ma2018} and Shu et al.\cite{Sh2021} have proposed asymptotic variance estimators for various causal estimands, with the latter focusing particularly on the Cox proportional hazard model. As mentioned in Shu et al.\cite{Sh2021}, the variance estimator for the ATE is conservative, while those for other estimands are not. Therefore, from a practical standpoint, it is also important to determine when CIs derived from these variance estimators are conservative.

%\vspace{0.5cm}
%\noindent
%{\bf Acknowledgements:} %We would like to express our gratitude to the editors for their useful comments, which have contributed greatly to focusing and improving the manuscript. We would like to thank Editage (www.editage.com) for the English language editing.
%The simulation analysis programs are available at the following URL:
%\begin{itemize}
%\item \url{https://github.com/SOrihara/###}

%\end{itemize}
%\vspace{0.5cm}
%\noindent
%{\bf Declaration of interest:}
%The authors declare they have no conflict of interest regarding this research study and paper.

%\vspace{0.5cm}
%\noindent
%{\bf Data availability statement:}
%Not applicable.

\newpage
\appendix
\section{Details of the IPW Estimator and the Asymptotic Property of the WATE}
In this manuscript, the propensity score is estimated as a MLE: $\sum_{i=1}^{n}\bld{X}_{i}\left(T_{i}-e_{i}(\bld{\alpha})\right)=\bld{0}$. To summarize the discussions in the main manuscript, both the IPW estimator and the propensity score estimator can be encapsulated by the following single estimating equation\cite{Ma2019}
\begin{align}
\label{EE1}
\sum_{i=1}^{n}\left(
\begin{array}{c}
\bld{X}_{i}\left(T_{i}-e_{i}(\bld{\alpha})\right)\\
W_{i}T_{i}(Y_{i}-\mu_{w1})\\
W_{i}(1-T_{i})(Y_{i}-\mu_{w0})
\end{array}
\right)=\sum_{i=1}^{n}\psi_{i}(\bld{\theta})=\bld{0},
\end{align}
where $\bld{\theta}:=\left(\bld{\alpha}^{\top},\mu_{w1},\mu_{w0}\right)^{\top}$, and  $\hat{\bld{\theta}}$ is the solution of (\ref{EE1}). Using the standard theory for the M-estimator, the asymptotic distribution of $\hat{\bld{\theta}}$ becomes
$$
\sqrt{n}\left(\hat{\bld{\theta}}-\bld{\theta}^{0}\right)\stackrel{L}{\to}N\left(\bld{0},{\rm E}\left[\frac{\partial \psi(\bld{\theta}^{0})}{\partial \bld{\theta}^{\top}}\right]^{-1}{\rm E}\left[\psi(\bld{\theta}^{0})^{\otimes 2}\right]{\rm E}\left[\frac{\partial \psi(\bld{\theta}^{0})^{\top}}{\partial \bld{\theta}}\right]^{-1}\right),
$$
where $\bld{\theta}^{0}$ represents the true value of $\boldsymbol{\theta}$, implying that the expectation of (\ref{EE1}) is uniquely satisfied. Also, applying the delta method,
\begin{align}
\label{asy1_}
\sqrt{n}\left(\hat{\tau}_{w}-\tau_{w}^{0}\right)\stackrel{L}{\to}N\left(\bld{0},\sigma^2\right),
\end{align}
where 
$$
\sigma^2=\bld{c}^{\top}{\rm E}\left[\frac{\partial \psi(\bld{\theta}^{0})}{\partial \bld{\theta}^{\top}}\right]^{-1}{\rm E}\left[\psi(\bld{\theta}^{0})^{\otimes 2}\right]{\rm E}\left[\frac{\partial \psi(\bld{\theta}^{0})^{\top}}{\partial \bld{\theta}}\right]^{-1}\bld{c},
$$
and $\bld{c}=(\bld{0}^{\top},1,-1)^{\top}$.

From here, we consider each component of the asymptotic variance of (\ref{asy1_}). Since
$$
{\rm E}\left[\frac{Tw(e)}{Te+(1-T)(1-e)}\right]={\rm E}\left[\frac{(1-T)w(e)}{Te+(1-T)(1-e)}\right]={\rm E}[w(e)],
$$
$$
\frac{\partial}{\partial \bld{\alpha}}\frac{Tw(e)}{Te+(1-T)(1-e)}(Y-\mu_{w1})=\frac{T(w'(e)e-w(e))}{e}(1-e)(Y-\mu_{w1})\bld{X},
$$
and
$$
\frac{\partial}{\partial \bld{\alpha}}\frac{(1-T)w(e)}{Te+(1-T)(1-e)}(Y-\mu_{w0})=\frac{(1-T)(w'(e)(1-e)+w(e))}{1-e}e(Y-\mu_{w0})\bld{X},
$$
\begin{align}
\label{mat1}
{\rm E}\left[\frac{\partial \psi(\bld{\theta}^{0})}{\partial \bld{\theta}^{\top}}\right]=\left(
\begin{array}{ccc}
-{\rm E}\left[e(1-e)\bld{X}^{\otimes 2}\right]&\bld{0}&\bld{0}\\
{\rm E}\left[(w'(e)e-w(e))(Y_{1}-\mu_{w1})(1-e)\bld{X}^{\top}\right]&-{\rm E}[w(e)]&0\\
{\rm E}\left[(w'(e)(1-e)+w(e))(Y_{0}-\mu_{w10})e\bld{X}^{\top}\right]&0&-{\rm E}[w(e)]
\end{array}
\right),
\end{align}
where $A^{\otimes 2}=AA^{\top}$. Also,
\begin{align}
\label{mat2}
{\rm E}\left[\psi(\bld{\theta}^{0})^{\otimes 2}\right]&=\left(
\begin{array}{ccc}
{\rm E}\left[e(1-e)\bld{X}^{\otimes 2}\right]&&\\
{\rm E}\left[w(e)(Y_{1}-\mu_{w1})(1-e)\bld{X}^{\top}\right]&{\rm E}\left[\frac{w(e)^2(Y_{1}-\mu_{w1})^2}{e}\right]&\\
-{\rm E}\left[w(e)(Y_{0}-\mu_{w0})e\bld{X}^{\top}\right]&0&{\rm E}\left[\frac{w(e)^2(Y_{0}-\mu_{w0})^2}{1-e}\right]
\end{array}
\right).
\end{align}
Regarding (\ref{mat1}) and (\ref{mat2}), the symbols used in the subsequent discussions are
$$
(\ref{mat1})=\left(
\begin{array}{cc}
-A_{11}&O^{\top}\\
A_{12}&a_{22}{\rm I}
\end{array}
\right),\ \ \ \ \ (\ref{mat2})=\left(
\begin{array}{cc}
A_{11}&B_{12}^{\top}\\
B_{12}&B_{22}
\end{array}
\right),
$$
where ${\rm I}$ represents the identity matrix. Using the relationship
$$
A_{12}=-B_{12}+\left(
\begin{array}{c}
{\rm E}\left[w'(e)(Y_{1}-\mu_{w1})e(1-e)\bld{X^{\top}}\right]\\
{\rm E}\left[w'(e)(Y_{0}-\mu_{w0})e(1-e)\bld{X^{\top}}\right]
\end{array}
\right)=-B_{12}+\delta=:-\left(
\begin{array}{c}
b_{1}^{\top}\\
b_{2}^{\top}
\end{array}
\right)+\left(
\begin{array}{c}
\delta_{1}^{\top}\\
\delta_{2}^{\top}
\end{array}
\right),
$$
the asymptotic variance of (\ref{asy1_}) becomes
$$
\sigma^{2}=\frac{1}{a_{22}^2}(1,-1)\left(B_{22}+\delta A_{11}^{-1}\delta^{\top}-B_{12}A_{11}^{-1}B_{12}^{\top}\right)\left(
\begin{array}{c}
1\\
-1
\end{array}
\right).
$$
Therefore, the result of the main manuscript (\ref{asy2}) is obtained.

\section{Proof of Theorem 2}
In the former arguments, it can be simply calculated that $\delta_{2}-\delta_{1}=O$. Given that ${\rm E}\left[e(1-e)\boldsymbol{X}^{\otimes 2}\right]>O$, the statement is proven.

In the latter arguments, from the Schwartz inequality, 
$$
(\delta_{2}-\delta_{1})^{\top}A_{11}^{-1}\alpha\leq|(\delta_{2}-\delta_{1})^{\top}A_{11}^{-1}\alpha|\leq\left\{(\delta_{2}-\delta_{1})^{\top}A_{11}^{-1}(\delta_{2}-\delta_{1})\alpha^{\top}A_{11}^{-1}\alpha\right\}^{\frac{1}{2}}.
$$
Therefore, 
\begin{align}
\label{ineq_att2}
-\alpha^{\top}A_{11}^{-1}\alpha+2(\delta_{2}-\delta_{1})^{\top}A_{11}^{-1}\alpha&\leq-\alpha^{\top}A_{11}^{-1}\alpha+2\left\{(\delta_{2}-\delta_{1})^{\top}A_{11}^{-1}(\delta_{2}-\delta_{1})\alpha^{\top}A_{11}^{-1}\alpha\right\}^{\frac{1}{2}}\nonumber\\
&=\left\{\alpha^{\top}A_{11}^{-1}\alpha\right\}^{\frac{1}{2}}\left[-\left\{\alpha^{\top}A_{11}^{-1}\alpha\right\}^{\frac{1}{2}}+2\left\{(\delta_{2}-\delta_{1})^{\top}A_{11}^{-1}(\delta_{2}-\delta_{1})\right\}^{\frac{1}{2}}\right]
\end{align}
Since $\alpha^{\top}A_{11}^{-1}\alpha > 0$, the sufficient condition for (\ref{ineq_att2}) to become negative is for the term within $[\cdot]$ to be negative:
$$
4(\delta_{2}-\delta_{1})^{\top}A_{11}^{-1}(\delta_{2}-\delta_{1})<\alpha^{\top}A_{11}^{-1}\alpha.
$$
We specifically calculate $\delta_{2}-\delta_{1}$ as follows:
$$
\delta_{2}-\delta_{1}={\rm E}\left[(Y_{0}-Y_{1}-(\mu_{w0}-\mu_{w1}))e(1-e)\bld{X}\right].
$$
When considering $NN$, $\mu_{w0}-\mu_{w1}=0$. Therefore, since $Y_{0}-Y_{1}<1$, the statement is verified. In other cases, given that $Y_{0}-Y_{1}-(\mu_{w0}-\mu_{w1})<2$, the statement is similarly verified.

\section{Continuous Outcomes Situations}
In this section, we assume the following linear outcome models
\begin{align}
\label{out1}
Y_{t}=\beta_{0t}'+{\bld X}^{\top}\bld{\beta}_{xt}+\varepsilon_{t},
\end{align}
where ${\rm E}[\varepsilon_{t}]=0$, $Var(\varepsilon_{t})<\infty$, and $\varepsilon_{t}\indep(T,\bld{X},Y)$ ($t=0,1$). Based on this assumption, the WATE is expressed as
$$
\tau_{w}=\mu_{w1}-\mu_{w0}=\beta_{01}'-\beta_{00}'+\frac{{\rm E}\left[w(e)\bld{X}^{\top}\right]}{{\rm E}[w(e)]}\left(\bld{\beta}_{x1}-\bld{\beta}_{x0}\right).
$$
In subsequent discussions, we will employ the following reparametrization to simplify the discussion: $\mu_{wt}=\beta_{0t}'$. This reparametrization is consistent with the centering of the outcome models (\ref{out1}) with respect to the covariates
$$
Y_{t}=\mu_{wt}+\left({\bld X}-\frac{{\rm E}\left[w(e)\bld{X}\right]}{{\rm E}[w(e)]}\right)^{\top}\bld{\beta}_{xt}+\varepsilon_{t}=\mu_{wt}+{\bld{X}'}^{\top}\bld{\beta}_{xt}+\varepsilon_{t}.
$$
Note that when treatment effects are homogeneous (ie., $\bld{\beta}_{x1}=\bld{\beta}_{x0}$), $\tau_{w}=\beta_{01}'-\beta_{00}'$ for all estimands. Note also that the centering does not affect the estimated propensity score since the effect is absorbed into the intercept term. Under the potential outcome models (\ref{out1}), the observed outcome model becomes
$$
Y=TY_{1}+(1-T)Y_{0}=\mu_{w0}+T(\mu_{w1}-\mu_{w0})+{\bld{X}'}^{\top}\bld{\beta}_{x0}+T{\bld{X}'}^{\top}\left(\bld{\beta}_{x1}-\bld{\beta}_{x0}\right)+\varepsilon
$$

First, we consider the asymptotic variance of the ATT. The values of $\delta$ and $B_{22}$ become the following, respectively:
\begin{align*}
\delta&=\left(
\begin{array}{c}
\delta_{1}^{\top}\\
\delta_{2}^{\top}
\end{array}
\right)=\left(
\begin{array}{c}
{\rm E}\left[(Y_{1}-\mu_{w1})e(1-e)\bld{X^{\top}}\right]\\
{\rm E}\left[(Y_{0}-\mu_{w0})e(1-e)\bld{X^{\top}}\right]
\end{array}
\right)=\left(
\begin{array}{c}
\bld{\beta}_{x1}^{\top}{\rm E}\left[e(1-e)\bld{X}^{\otimes 2}\right]\\
\bld{\beta}_{x0}^{\top}{\rm E}\left[e(1-e)\bld{X}^{\otimes 2}\right]
\end{array}
\right),\\
B_{12}&=\left(
\begin{array}{c}
b_{1}^{\top}\\
b_{2}^{\top}
\end{array}
\right)=\left(
\begin{array}{c}
{\rm E}\left[(Y_{1}-\mu_{w1})e(1-e)\bld{X}^{\top}\right]\\
-{\rm E}\left[(Y_{0}-\mu_{w0})e^2\bld{X}^{\top}\right]
\end{array}
\right)=\left(
\begin{array}{c}
\bld{\beta}_{x1}^{\top}{\rm E}\left[e(1-e)\bld{X}^{\otimes 2}\right]\\
-\bld{\beta}_{x0}^{\top}{\rm E}\left[e^2\bld{X}^{\otimes 2}\right]
\end{array}
\right).
\end{align*}
Since $b_{1}=\delta_{1}$ and $b_{2}=\delta_{2}-\bld{\beta}_{x0}^{\top}{\rm E}\left[e\bld{X}^{\otimes 2}\right]$, the second and third term of (\ref{asy2}) become
\begin{align*}
(1,-1)\left(\delta A_{11}^{-1}\delta^{\top}-B_{12}A_{11}^{-1}B_{12}^{\top}\right)\left(
\begin{array}{c}
1\\
-1
\end{array}
\right)&\\
&\hspace{-6.5cm}=\delta_{1}^{\top}A_{11}^{-1}\delta_{1}-2\delta_{1}^{\top}A_{11}^{-1}\delta_{2}+\delta_{2}^{\top}A_{11}^{-1}\delta_{2}-b_{1}^{\top}A_{11}^{-1}b_{1}+2b_{1}^{\top}A_{11}^{-1}b_{2}-b_{2}^{\top}A_{11}^{-1}b_{2}\\
&\hspace{-6.5cm}=-2\bld{\beta}_{x1}^{\top}{\rm E}\left[e(1-e)\bld{X}^{\otimes 2}\right]A_{11}^{-1}{\rm E}\left[e\bld{X}^{\otimes 2}\right]\bld{\beta}_{x0}+2\bld{\beta}_{x0}^{\top}{\rm E}\left[e\bld{X}^{\otimes 2}\right]A_{11}^{-1}{\rm E}\left[e(1-e)\bld{X}^{\otimes 2}\right]\bld{\beta}_{x0}\\
&\hspace{-6cm}-\bld{\beta}_{x0}^{\top}{\rm E}\left[e\bld{X}^{\otimes 2}\right]A_{11}^{-1}{\rm E}\left[e\bld{X}^{\otimes 2}\right]\bld{\beta}_{x0}\\
&\hspace{-6.5cm}=2\bld{\beta}_{x0}^{\top}{\rm E}\left[e\bld{X}^{\otimes 2}\right]\left(\bld{\beta}_{x0}-\bld{\beta}_{x1}\right)-\bld{\beta}_{x0}^{\top}{\rm E}\left[e\bld{X}^{\otimes 2}\right]A_{11}^{-1}{\rm E}\left[e\bld{X}^{\otimes 2}\right]\bld{\beta}_{x0}
\end{align*}
From the result, the following relationship can be proved.
\begin{theo}$\phantom{a}$\\
For the ATT, when treatment effects are homogeneous (ie., $\bld{\beta}_{x1}=\bld{\beta}_{x0}$), the second and third terms of (\ref{asy2}) are precisely negative. When there is only constant heterogeneity (ie., $\bld{\beta}_{x1}=\gamma\bld{\beta}_{x0}$), a sufficient condition for the second and third terms of (\ref{asy2}) to be precisely negative is the existence of a value $\gamma\in\mathbb{R}$ such that
$$
{\rm E}\left[e(1-e)\bld{X}^{\otimes 2}\right]^{-1}{\rm E}\left[e\bld{X}^{\otimes 2}\right]-2(1-\gamma){\rm I}>O.
$$
This is clearly satisfied when $\gamma\geq1$.
\end{theo}\noindent
\begin{proof}
Since the former statement is obvious, we will only prove the latter statement. When $\bld{\beta}_{x1}=\gamma\bld{\beta}_{x0}$,
\begin{align*}
2(1-\gamma)\bld{\beta}_{x0}^{\top}{\rm E}\left[e\bld{X}^{\otimes 2}\right]\bld{\beta}_{x0}-\bld{\beta}_{x0}^{\top}{\rm E}\left[e\bld{X}^{\otimes 2}\right]A_{11}^{-1}{\rm E}\left[e\bld{X}^{\otimes 2}\right]\bld{\beta}_{x0}&\\
&\hspace{-10cm}=\bld{\beta}_{x0}^{\top}{\rm E}\left[e\bld{X}^{\otimes 2}\right]A_{11}^{-1}\left\{2(1-\gamma){\rm E}\left[e(1-e)\bld{X}^{\otimes 2}\right]-{\rm E}\left[e\bld{X}^{\otimes 2}\right]\right\}\bld{\beta}_{x0}
\end{align*}
By focusing on the term within $\{\cdot\}$, the statement can be derived.
\end{proof}\noindent
This theorem suggests that the simple CI yields a conservative CI when there is a homogeneous treatment effect, and we are interested in the ATT. Additionally, if the interaction effect between the treatment and all confounders is proportional to $\gamma$, the simple CI also yields a conservative CI when the interaction effect is superior to the effect from covariates alone. This scenario arises when the risk factors of interest exhibit significant interaction effects with a treatment.

Next, we consider the asymptotic variance of the ATO. In the same manner as the ATT, the values of $\delta$ and $B_{22}$ become the following, respectively:
\begin{align*}
\delta&=\left(
\begin{array}{c}
\delta_{1}^{\top}\\
\delta_{2}^{\top}
\end{array}
\right)=\left(
\begin{array}{c}
{\rm E}\left[(Y_{1}-\mu_{w1})e(1-e)(1-2e)\bld{X^{\top}}\right]\\
{\rm E}\left[(Y_{0}-\mu_{w0})e(1-e)(1-2e)\bld{X^{\top}}\right]
\end{array}
\right)=\left(
\begin{array}{c}
\bld{\beta}_{x1}^{\top}{\rm E}\left[e(1-e)(1-2e)\bld{X}^{\otimes 2}\right]\\
\bld{\beta}_{x0}^{\top}{\rm E}\left[e(1-e)(1-2e)\bld{X}^{\otimes 2}\right]
\end{array}
\right),\\
B_{12}&=\left(
\begin{array}{c}
b_{1}^{\top}\\
b_{2}^{\top}
\end{array}
\right)=\left(
\begin{array}{c}
{\rm E}\left[(Y_{1}-\mu_{w1})e(1-e)^2\bld{X}^{\top}\right]\\
-{\rm E}\left[(Y_{0}-\mu_{w0})e^2(1-e)\bld{X}^{\top}\right]
\end{array}
\right)=\left(
\begin{array}{c}
\bld{\beta}_{x1}^{\top}{\rm E}\left[e(1-e)^2\bld{X}^{\otimes 2}\right]\\
-\bld{\beta}_{x0}^{\top}{\rm E}\left[e^2(1-e)\bld{X}^{\otimes 2}\right]
\end{array}
\right).
\end{align*}
Through the similar calculation as the ATT (\ref{ATTsnd}), the second and third term of (\ref{asy2}) become
\begin{align*}
(1,-1)\left(\delta A_{11}^{-1}\delta^{\top}-B_{12}A_{11}^{-1}B_{12}^{\top}\right)\left(
\begin{array}{c}
1\\
-1
\end{array}
\right)&\nonumber\\
&\hspace{-6.5cm}=\left(\bld{\beta}_{x0}-2\bld{\beta}_{x1}\right)^{\top}{\rm E}\left[e(1-e)\bld{X}^{\otimes 2}\right]\bld{\beta}_{x0}-2\left(\bld{\beta}_{x1}-2\bld{\beta}_{x0}\right)^{\top}{\rm E}\left[e^2(1-e)\bld{X}^{\otimes 2}\right]\left(\bld{\beta}_{x1}-\bld{\beta}_{x0}\right)\\
&\hspace{-6cm}+3\left(\bld{\beta}_{x1}-\bld{\beta}_{x0}\right)^{\top}{\rm E}\left[e^2(1-e)\bld{X}^{\otimes 2}\right]{\rm E}\left[e(1-e)\bld{X}^{\otimes 2}\right]^{-1}{\rm E}\left[e^2(1-e)\bld{X}^{\otimes 2}\right]\left(\bld{\beta}_{x1}-\bld{\beta}_{x0}\right)
\end{align*}
From the result, the following relationship can be proved.
\begin{theo}$\phantom{a}$\\
For the ATO, when treatment effects are homogeneous (ie., $\bld{\beta}_{x1}=\bld{\beta}_{x0}$), the second and third terms of (\ref{asy2}) are precisely negative. When there is only constant heterogeneity (ie., $\bld{\beta}_{x1}=\gamma\bld{\beta}_{x0}$), sufficients condition for the second and third terms of (\ref{asy2}) to be precisely negative is the existence of a value $\gamma>1$ such that
\begin{align}
\label{ATOcond2}
\frac{2(\gamma-2)}{3(\gamma-1)}{\rm I}-{\rm E}\left[e(1-e)\bld{X}^{\otimes 2}\right]^{-1}{\rm E}\left[e^2(1-e)\bld{X}^{\otimes 2}\right]>O,
\end{align}
\end{theo}\noindent
\begin{proof}
Since the former statement is obvious, we will only prove the latter statement. When $\bld{\beta}_{x1}=\gamma\bld{\beta}_{x0}$,
\begin{align*}
(1-2\gamma)\bld{\beta}_{x0}^{\top}{\rm E}\left[e(1-e)\bld{X}^{\otimes 2}\right]\bld{\beta}_{x0}-2(\gamma-1)(\gamma-2)\bld{\beta}_{x0}^{\top}{\rm E}\left[e^2(1-e)\bld{X}^{\otimes 2}\right]\bld{\beta}_{x0}&\\
&\hspace{-13cm}+3(\gamma-1)^2\bld{\beta}_{x0}^{\top}{\rm E}\left[e^2(1-e)\bld{X}^{\otimes 2}\right]{\rm E}\left[e(1-e)\bld{X}^{\otimes 2}\right]^{-1}{\rm E}\left[e^2(1-e)\bld{X}^{\otimes 2}\right]\bld{\beta}_{x0}\\
&\hspace{-13.5cm}=(1-2\gamma)\bld{\beta}_{x0}^{\top}{\rm E}\left[e(1-e)\bld{X}^{\otimes 2}\right]\bld{\beta}_{x0}+(\gamma-1)\bld{\beta}_{x0}^{\top}{\rm E}\left[e^2(1-e)\bld{X}^{\otimes 2}\right]{\rm E}\left[e(1-e)\bld{X}^{\otimes 2}\right]^{-1}\\
&\hspace{-13cm}\times\left\{-2(\gamma-2){\rm E}\left[e(1-e)\bld{X}^{\otimes 2}\right]+3(\gamma-1){\rm E}\left[e^2(1-e)\bld{X}^{\otimes 2}\right]\right\}\bld{\beta}_{x0}
\end{align*}
When $\gamma\geq0.5$, the first term is negative definite. When $\gamma>1$, the second term is negative definite under the condition (\ref{ATOcond2}). When $1>\gamma\geq0.5$ the second term is negative definite under the condition
$$
\frac{2(\gamma-2)}{3(\gamma-1)}{\rm I}-{\rm E}\left[e(1-e)\bld{X}^{\otimes 2}\right]^{-1}{\rm E}\left[e^2(1-e)\bld{X}^{\otimes 2}\right]<O.
$$
However, from the relationship
\begin{align}
\label{posi}
{\rm E}\left[e^2(1-e)\bld{X}^{\otimes 2}\right]={\rm Pr}(T=1){\rm E}\left[e(1-e)\bld{X}^{\otimes 2}|T=1\right]<{\rm E}\left[e(1-e)\bld{X}^{\otimes 2}\right],
\end{align}
There are no situations where the condition can be satisfied.
\end{proof}\noindent
Note that when $\gamma<0.5$, simple sufficient conditions cannot be derived because the first term of the above formula is positive definite. This implies that when there's a small, or negative interaction effect between the treatment and all confounders, the simple CI may not yield a conservative CI. When $\gamma>1$, relatively straightforward condition (\ref{ATOcond2}) can be derived; however, interpretation remains challenging.

To address this, we use the following relationship (\ref{posi}). When ${\rm Pr}(T=1)\approx0$, it is expected that ${\rm E}\left[e(1-e)\bld{X}^{\otimes 2}\right]^{-1}{\rm E}\left[e^2(1-e)\bld{X}^{\otimes 2}\right]\approx O$. In this situation (\ref{ATOcond2}) under $2\geq\gamma>1$ is not hold. Whereas, (\ref{ATOcond2}) under $\gamma>2$ is hold. This means that if there are only a few members in the treatment group, and there are large interaction effects between the treatment and all confounders, the simple CI yields a conservative CI. When ${\rm Pr}(T=1)\approx1$, it is expected that ${\rm E}\left[e(1-e)\bld{X}^{\otimes 2}\right]^{-1}{\rm E}\left[e^2(1-e)\bld{X}^{\otimes 2}\right]\approx {\rm I}$. In this situation,
$$
\left(\frac{2(\gamma-2)}{3(\gamma-1)}-1\right){\rm I}=\frac{-1-\gamma}{3(\gamma-1)}{\rm I}.
$$
Obviously, (\ref{ATOcond2}) is not satisfied. Note that when ${\rm Pr}(T=1)\approx0$ and $\gamma>0.5$, (\ref{ATOcond2}) may not always hold from the above discussions; however, the first term of (\ref{asy2}) dominates. Therefore, the simple CI also yields a conservative CI in this situation.

Summarizing the discussions above, for the ATO, there is no universal scenario where the simple CI is applicable; the exact CI is more appropriate. However, in certain situations, such as when there are homogeneous treatment effects, when there exists significant treatment-confounder interactions, or when there are many members in the control groups, the simple CI might work effectively.
\end{document}